\documentclass[runningheads]{llncs}
\usepackage{graphicx}
\usepackage{psfig}

\begin{document}

\pagestyle{headings}

\mainmatter

\title{Local Information Based Algorithms for Packet\\
Transport in Complex Networks}

\titlerunning{Lecture Notes in Computer Science}

\author{Bernard Kujawski\inst{1}
\and G.J. Rodgers\inst{1} \and Bosiljka Tadi\'{c} \inst{2} }

\authorrunning{Bernard Kujawski et al.}

\institute{Department of Mathematical Sciences; Brunel University, Uxbridge;\\
Middlesex UB8 3PH; UK\\
\email{\{bernard.kujawski,g.j.rodgers\}@brunel.ac.uk}\\
\and
Department for Theoretical Physics; Jo\v{z}ef Stefan Institute, P.O. Box 3000;\\
SI-1001 Ljubljana; Slovenia\\
\email{bosiljka.tadic@ijs.si}}

\maketitle

\begin{abstract}

{\bf} We introduce four algorithms for packet transport in complex
networks. These algorithms use deterministic rules which depend, in
different ways, on the degree of the node, the number of packets
posted down each edge, the mean delivery time of packets sent down
each edge to each destination and the time since an edge last
transmitted a packet. On scale-free networks all our algorithms are
considerably more efficient and can handle a larger load  than the
random walk algorithm. We consider in detail various attributes of
our algorithms, for instance we show that an algorithm that bases
its decisions on the mean delivery time jams unless it incorporates
information about the degree of the destination node.
\end{abstract}

\section{Introduction}

Complex networks can be used to model a wide range of physical and
technological systems. One of the most interesting dynamical
problems on network is transport, which can give us some insight
into the transport of information in technology based communication
networks like the internet \cite{faloutsos}, the World Wide Web
\cite{albert},\cite{Huberman} or phone call networks \cite{Adamic}.
Here we use the term \emph{transport} to mean transport of particles, which
are packets in a network. Thus our model falls within the Network
Layer of the OSI Reference Model and the algorithms described in
section $3$ are routing algorithms that belong to the Network Layer
of the OSI Reference Model. Of particular interest is the phenomenon
of load in a network, as a function of the rate of packet creation
R, which has been investigated for models of communication networks
\cite{Tadic},\cite{Arenas},\cite{Sole} and in real networks
\cite{Jacobson}.

Typically the problem of transport is investigated using either a random
walk algorithm \cite{Tadic}, or the shortest path algorithm used by
most internet protocols. The difficulty with these approaches is
that random walk algorithm is very inefficient for transport in
technology based communication networks and shortest path algorithm
requires, for its implementation, information about all connections
in network. In this paper we focus on algorithms that use local
information about the topology, along with information about the flux of
packets between neighbors, the link load and the time taken to deliver
packets. We propose four algorithms that use some or all of these
properties to deliver packets in a network.

In section 2 we describe the algorithm that we use to perform numerical
simulations of our models. In section 3 we discuss the algorithms that
packets use to find their destinations and in section 4 we show our results.
In section 5 we summarise our results.

\section{The Program}
A program was written to simulate packet transport on a network that
does not depend on the size of the network or its
topology. At the beginning of the program an external file with
the adjacency matrix of the network is read in.
We focus on the internet and
consequently we treat nodes in our network as if they were routers.
The connections between the routers
have the same capacity for all networks. Such a model can
not only be used to model internet packet transport but also for a range of
transport networks in which the nodes have local routing information.

\textbf{Each node}:
\begin{itemize}
  \item Generates a new packet with probability $r=R/N$ and
  with a randomly chosen destination, where $R$ is a fixed rate
  for the whole network, and $N$ is the number of nodes in network.
  \item Stores packets in a queue, which has maximum length
  is $L=1000$. Packets are despatched from the queue in a first
  in first out (FIFO) order.
  \item Sends packets to its neighbours.
\end{itemize}

\textbf{Each node has information about}:
\begin{itemize}
  \item The address of all its
neighbours (they have unique indices $j$).
  \item The degree of its neighbours - $k(i)$.
  \item Flow through all its neighbours, which is measured by
\begin{itemize}
  \item The number of packets posted down each edge to neighbour $i$ - the Link Load -  $C(i)$.
  \item  The number of packets sends through neighbour $i$, which have reached their destination
  - $N_P(i)$.
  \item The sum of the delivery times of all the packets sent through neighbour $i$ that have reached their
destination - $T_P(i)$.
\item The time interval since an edge last
transmitted a packet to neighbour $i$ and current time step -
$\Delta T(i)$.
\end{itemize}

\end{itemize}

The index $i$ enumerates each neighbour of node $k$ and each node
keeps all the statistics about its neighbours. Quantities $C(i)$,
$N_P(i)$, $T_P(i)$ and $\Delta T(i)$ describe node $i$ from the
perspective of node $k$. Each node is described by its neighbours
and all properties can be different for all neighbours that describe
node $i$.

The initialization part of the program sets up the network topology,
the nodes and all the tables used by them. Inside the main loop a
time step is incremented, and within that a loop over all nodes
calculates and updates the statistics. The loop over all nodes
includes three basics routines, which are run for each node;
generating new packets, checking its queue for packets with its
address and sending packets to its neighbours. Each node generates a
packet with a randomly chosen destination with probability $R/N$.
The node checks its own queue for packets addressed to itself. When
it finds one of these it deletes it from the queue and updates the
statistics $N_P(i)$ and $T_P(i)$ for all the nodes on the packet's
path. Each packet keeps track of its own path. The node sends
packets to its neighbours by taking the first packet in its queue
and checking the packet destination address. If the packet is
addressed to one of its neighbour, the node will send it to the
neighbour. If it is not, the node will use the \emph{algorithm} to
find where to send the packet. During this posting step the $C(i)$
property is updated. When node $k$ sends packets to node $i$, the
number of sent packets $C(i)$ increases. After this loop over all
the nodes is completed the quantities $\Delta T(i)$ and the mean
delivery time of packets sent down each edge $N_P(i)/T_P(i)$ are
updated for all nodes.

\section{Algorithms}
The most important element in transport is the rule that determines
the direction in which a packet is sent. A transport network without
a rule is a random walk network. We call this rule the
\emph{algorithm}. It describes how nodes deal with packets and
should help packets to get to their destination. Not all algorithms
help packets to reach destinations, poor algorithms can easily be
worse than the random walk algorithm. All algorithms considered in
this paper work with`deterministic rules.

The \emph{shortest time}(ST) algorithm is our basic algorithm that
uses information about the mean delivery time $T_P(i)/ N_P(i)$ and
the time interval between the last packet that came to node $i$ and
actual time step. The ST algorithm finds the minimum value
\begin{equation}\label{ST}
  S_k=\min\left[\frac{T_P(i)}{N_P(i)}\frac{1}{\Delta
  T(i)}\right]_{i=1\ldots n}
\end{equation}
in order to determine which node to send the packet to. The idea of
this algorithm is to try and find the minimum travel time for each
packet between source and destination. At the start of the
simulation $S$ is equal to $0$ for all neighbours.  Because the
update of $T_P(i)/ N_P(i)$ only occurs when a packet arrives at its
destination, it can take a number of time steps before $T_P(i)/
N_P(i)$ becomes non-zero. The inclusion of the reciprocal of $\Delta
T(i)$ in $S$ ensures that the algorithm does not get into a state
where it never sends a packet down certain links which have a large
mean delivery time. This state is particularly likely to occur at
the start of the simulation. The inclusion of the reciprocal of
$\Delta T(i)$ in $S$ also prevents overcrowding when a node finds a
node which is clearly better than all its other neighbours. Hence,
because of the inclusion of $\Delta T(i)$ more nodes take part in
the transport and in this way the large node do not become
overcrowded. Because the algorithm with $T_P(i)/ N_P(i)$ is looking
for minimum delivery time we call it the \emph{shortest time} (ST)
algorithm. To start this algorithm, and the STD algorithm, which we
will introduce shortly, we use the random walk algorithm. We only
use the deterministic algorithms at a node when all the values of
$S$ of its neighbours are greater than 0. Without this initial
random walk procedure both the ST and the STD algorithms would jam
almost immediately. The \emph{shortest time and degree} (STD)
algorithm is a modification of the ST algorithm. It uses information
about the local topology, the degree. This helps packets avoid the
nodes with the largest degree, which are mostly overcrowded. The
idea of incorporating information about the degree of nodes in the
transport algorithm was discussed in \cite{Yan} and \cite{Wang}. In
these papers models were introduced in which nodes were selected at
a rate proportional to a power of their degree. It was found that
the most efficient algorithm was one in which the the probability of
selecting a node of degree $k$ was proportional to  $1/k$ \cite{Yan}
and \cite{Wang}. The STD algorithm is defined by
\begin{equation}\label{STD}
  S_k=\min\left[\frac{T_P(i)}{N_P(i)}\frac{1}{\Delta
  T(i)}k(i)\right]_{i=1\ldots n}
\end{equation}
where $k(i)$ is a degree of node $i$ and $k(i)>1$. This last
assumption allows the algorithm to avoid dead-end nodes. A node with
degree $k=1$ can only receive a packet that is addressed to itself.
The STD algorithm uses both temporal properties and also information
about the local connectivity. For transport in a scale-free network
the most important nodes are those with the largest degree. But
because their neighbours send these nodes a large number of packets
the queues at these nodes can become overcrowded. Information about
the degree helps the algorithm to avoid these nodes, but it does not
mean than they are not used.

The \emph{connections and degree} (CD) algorithm and the
\emph{connections, degree and shortest time} (CDT) algorithm use
information about the link load $C(i)$. Because of this the random
walk starting procedure used in the ST and STD algorithms is not
required for the CD and CDT algorithms. The CD algorithm uses only
information about the link load and the degree. The CD algorithm is
defined by
\begin{equation}\label{CD}
  S_k=\min[C(i)k(i)]_{i=1\ldots n}
\end{equation}
where $C(i)$ is a number of packets that node $k$ sends to node
$i$.

For this algorithm $S$ equals $0$ at the start, but $C(i)$ is
updated almost immediately. When node $k$ sends a packet then it
automatically increases the value of $C(i)$. There is no need to
wait for information from the destination about the delivery time
like in the ST and STD algorithms. In this way CD algorithm improves
very quickly and the random walk is not needed. The link load,
$C(i)$, quantity helps the algorithm to deliver packets and ensures
that almost all nodes take part in the transport. The degree
quantity helps to prevent the largest nodes from becoming
overcrowded. In this algorithm there is no property that can be
optimised, unlike in the ST and STD algorithms where the delivery
time is optimised.

The CDT algorithm is intermediate between the CD and the ST
algorithms. It optimises the delivery time and does not need the
random the walk starting procedure because it includes a dependance
on the link load, $C(i)$. The dependence on degree prevents large
nodes becoming overcrowded. For the CDT algorithm, the starting
procedure is the same as for the CD algorithm except that we set
\begin{equation}\label{TimeElem}
  \frac{T_P(i)}{N_P(i)}\frac{1}{\Delta
  T(i)}
\end{equation}
equal to $1$ at the start to avoid $0$ value. This means that we do not
need to start off with a random walk algorithm as in the ST and STD algorithms.
The CDT algorithm is defined by
\begin{equation}\label{CDT}
  S_k=\min\left[\frac{T_P(i)}{N_P(i)}\frac{1}{\Delta
  T(i)}C(i)k(i)\right]_{i=1\ldots n}\qquad\mbox{with }k(i)>1.
\end{equation}

We use the learning property to describe behavior of an algorithm in
the beginning. By learning we mean the proportion of links whose
value of $S$ has changed since $t=0$. The CD and CDT algorithms
learn the most quickly. After $5000$ time steps they tried $95\%$ of
links. This is because the link load, $C(i)$, changes when a packet
is sent down it whereas $T_P(i)/ N_P(i)$, used by the ST and STD
algorithms, only changes when a packet sent down it gets to its
destination. That is way the ST and STD algorithms need the random
walk starting procedure. With this procedure after $5000$ time steps
$35\%$ of links were tried. For the ST algorithm without the random
walk starting procedure it was $5\%$. The speed of learning is
important because when a network learns slowly, the network only
uses a small proportion of its links for transport over a long
period of time, which means that the network is easily jammed when a
region of the network becomes overcrowded.

\section{Results}
We consider transport on the Barabasi and Albert model of a network
\cite{Albert_base} with $N=1000$ nodes and $m=2$. The parameter $m$
is the number of links of a new node that is added to network. When
$m=2$ the network includes loops and has relative small number of
connections. Our research show that this network jams for lower
values of the posting rate than networks with $m=1$ or $m=3$ and
higher. In this work we use a posting rate of $R=0.1$. This means
that each node creates a packet with probability $R/N$. The number
of time steps for all our simulations is $500,000$. We present
results for the STD, CD and CDT algorithms. We do not consider the
ST algorithm any further because it isn't stable and always jams.

\begin{figure}
\centerline{\psfig{figure=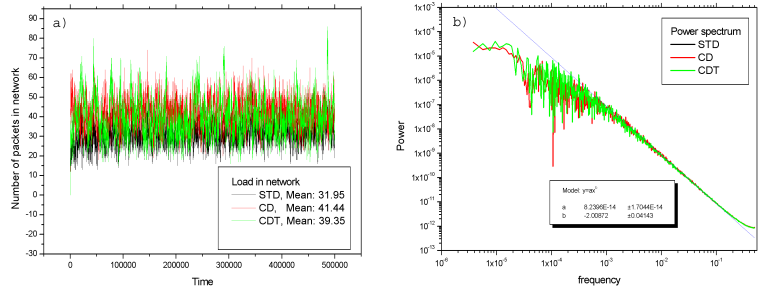,height=4cm}} \caption{The load
properties. $a)$ Load in the network for the STD,CD and CDT
algorithms. $b)$ The power spectrum for all algorithms is the same and shows
shows that noise in network is uncorrelated.} \label{LoadProp}
\end{figure}

In figure \ref{LoadProp}a we show the load in the network,
the number of packets that are still in the network.  All
three algorithms are stable. We compared the level of load by
finding the mean value of the number of packets in the network. The
best algorithm with smallest mean value is the STD algorithm. For
the CD and CDT the values are almost the same.

\begin{figure}[htb]
 \centering
 \centerline{\psfig{figure=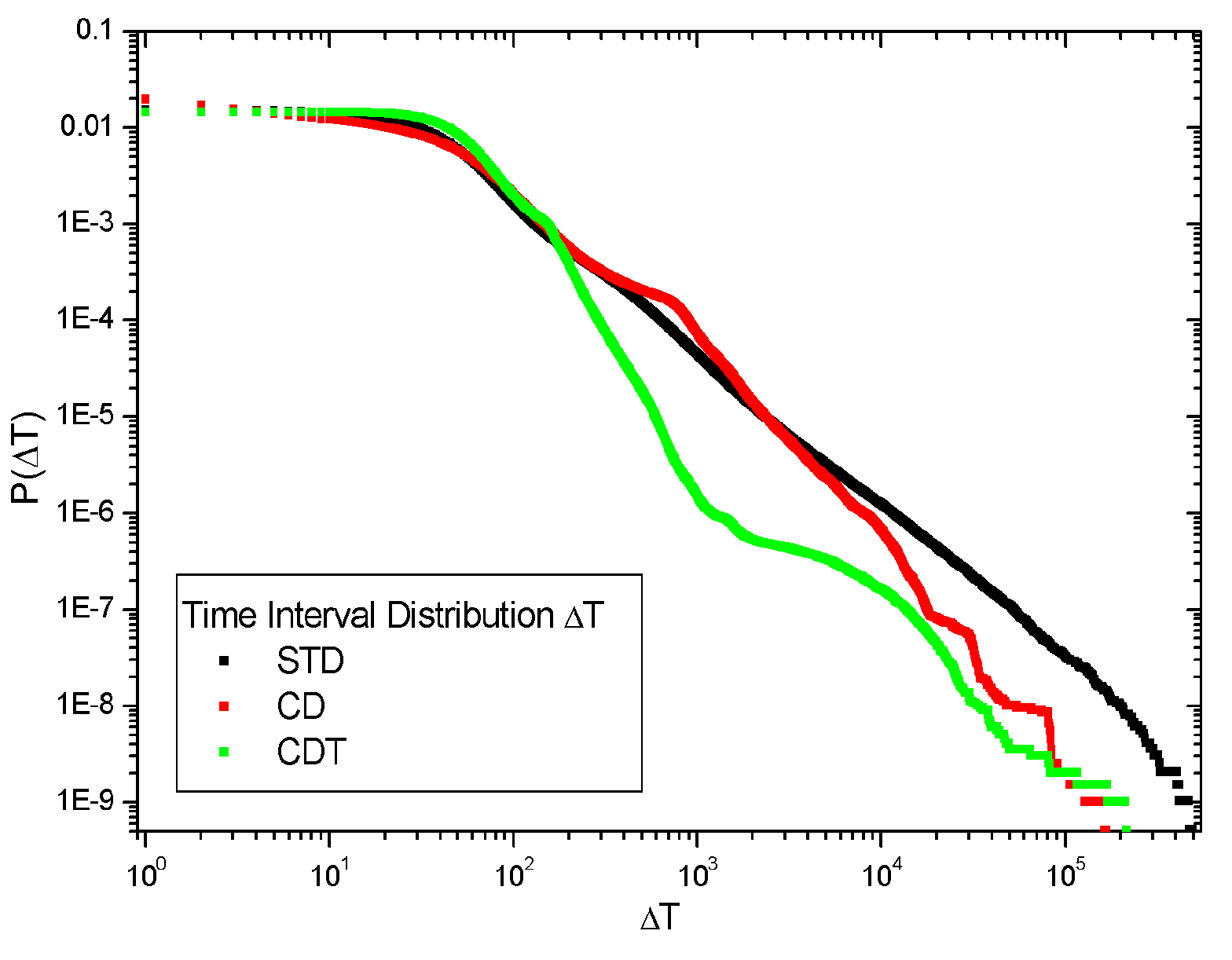,height=4cm}}
  \caption[TIDdt]{Distribution of time interval $\Delta T(i)$.}
  \label{TIDdt}
\end{figure}

The number of packets in network can be treated as a noise in the
network. Measuring the power spectrum of this noise shows that there
are correlations in the number of packets in network. For all our
algorithms the power spectrum (Fig.\ref{LoadProp}b) is the same and
the slope has $-2$.  It means that the noise in network is like
$1/f^2$; uncorrelated noise with short-range correlations only.
We measured the distribution of the time interval $\Delta T(i)$, the
time that nodes wait for packets, and the results are shown in
figure \ref{TIDdt}. This is an important quantity for the SDT and
CDT algorithms as without the $\Delta T(i)$ term these networks
easily jam. For the STD algorithm the distribution of $\Delta T(i)$
has a tail and on a double logarithmic scale has a slope $b=-3/2$.
The cut-off comes from the finite time of the simulation. The first
part of the distribution for all algorithms is flat. For the CDT
algorithm the function falls faster than for the STD. This is
connected with the inclusion of the link load in the CDT algorithm,
which means that more links are used and long time intervals of
$\Delta T(i)$ do not occur as frequently as in the STD algorithm.
The CD algorithm does not use $\Delta T(i)$ but we measured it to
compare it to the other models.

\begin{figure}
\centerline{\psfig{figure=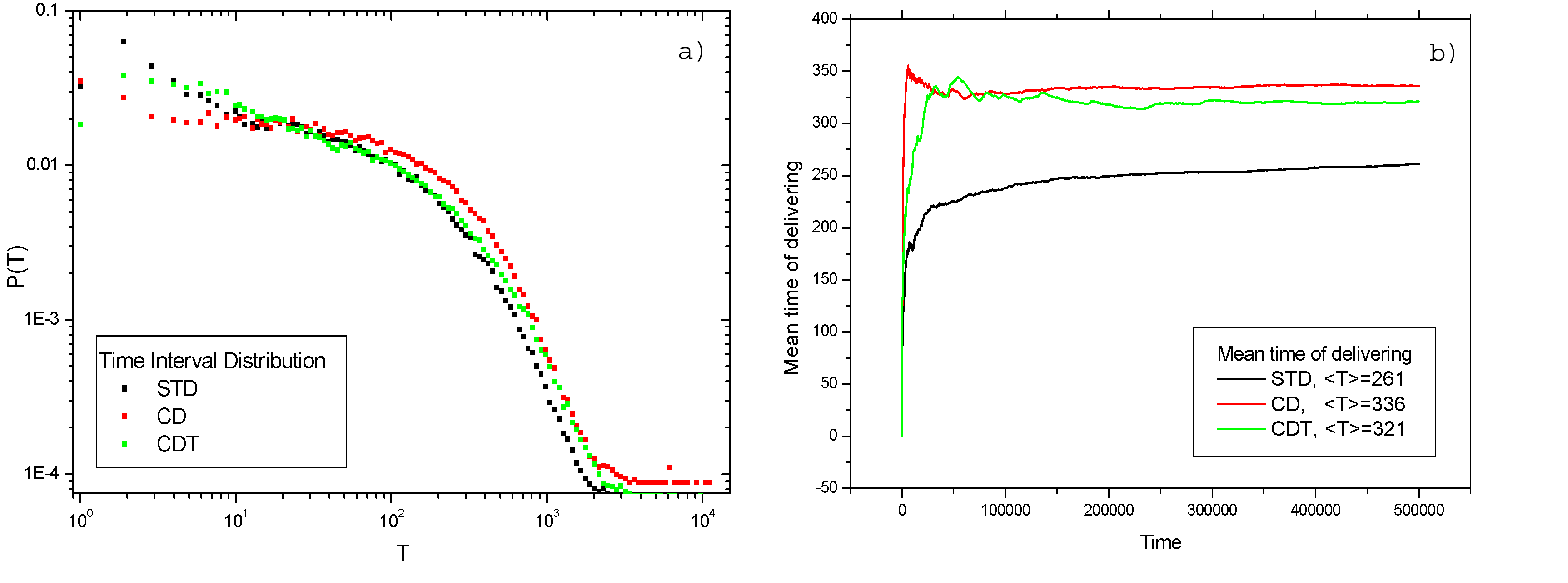,height=4cm}} \caption{The time
delivery quantities for the STD, CD and CDT algorithms. $a)$ The
time interval distribution for packets between a source and a
destination. $b)$ The mean delivery time.} \label{TimeProp}
\end{figure}

The distribution of packet delivery time (Fig.\ref{TimeProp}a) is
similar for all the algorithms. However the distribution shows that
the number of packets delivered in a short time is different for
each algorithm For the STD algorithm packets are delivered quickly
more frequently than for the CD and CDT algorithms. The STD
algorithm finds the paths with the shortest delivery time because,
whilest the CD and CDT algorithms are distributing the transport
across the network, because their algorithms use the link load
$C(i)$, the STD algorithm is looking for shortest delivery times.
The distribution for the CDT algorithm is intermediate between the
STD and CD algorithms because the CDT algorithm depends on the link
load $C(i)$ and the shortest time statistics.

The time series for the overall mean delivery time
(Fig.\ref{TimeProp}b) show that algorithms involving the statistics
for $T_P(i)/N_P(i)$ do not learn. The mean delivery time for CD and
CDT is almost the same. The algorithms reach a stable mean delivery
time and do not optimise it. Obviously for the CD algorithm no
optimisation is possible because there is no quantity that could be
optimized.

The result for the STD and CDT algorithms arise through two effects.
The first is the inclusion of $\Delta T(i)$ in $S$ that send packets
to rarely used links that often are not the best ones for transport.
On the other hand without $\Delta T(i)$ all the algorithms with the
mean time property start jamming. Secondly is the inclusion of the
degree in $S$, which means that algorithms prefer to send packets to
nodes with a small degree which makes the delivery time long.

\section{Conclusions}
The algorithms STD, CD and CDT work well; for the same network and
for the same value of $R$ the random walk algorithm jams, and these
algorithms do not. One might expect that including the mean delivery
time of packets sent to node $i$, $T_P(i)/N_P(i)$, in $S$ would
optimize the delivery time. This does not happen because of the
dependence of $S$ on the delivery time, link load and degree. But on
the other hand without dependence on these terms the algorithms
cannot work properly. This the case in the ST algorithm, which works
better than the random walk algorithm, but much worse than the other
algorithms. When the shortest time property is used in the
scale-free network it needs to be balanced be degree quantity. The
existence of nodes with large degrees causes traffic congestion for
the shortest time algorithm. Using an algorithm which depends on
local degree information but without dependence on the mean time (CD
algorithm) works correctly but an algorithm without local degree
dependence and with the mean time dependence (ST algorithm) jams
easily. The biggest problem in implementing the STD and CDT
algorithms is in finding accurate value for the edge dependent
properties. A node needs a lot of connections through one link to
find it proper time statistics. Because the mean delivery time is
very long, it takes a lot of time to set up the edges dependent
properties for all nodes. In particular, the algorithms that depend
on the time $\Delta T(i)$ and the degree $k(i)$ do not jam but the
cost is in learning and the mean delivery time. The inclusion of the
$\Delta T(i)$ quantity in $S$, avoids jamming but destroys the
learning behavior promoted by the inclusion of the mean time
property in $S$. The degree property helps the algorithm to avoid
nodes with large degree, and hence helps prevent overcrowding, but
it also results in long delivery times. Our results show that in
scale free networks we cannot avoid using nodes with large degree.

In future work, it may be possible to develop an algorithm that uses
information on the mean local delivery time to find the optimal path
for transport. One possible extension of this work would be to use
an algorithm that allows a number of packets to be sent to a node in
one time step, depending on the degree of the node. This is
realistic because normally routers can use all their outputs almost
in a parallel way. The biggest problem in networks is that nodes
with a very high degree can receive as many packets as they have
inputs in one time step, but they usually send only one packet. When
we allow them to use all their outputs in one time step then jamming
will disappear.

\end{document}